\documentclass{SciPost}

\usepackage{hyperref}
\hypersetup{
    colorlinks,
    linkcolor={red!50!black}, 
    citecolor={blue!50!black},
    urlcolor={blue!80!black}
}
\usepackage[normalem]{ulem}
\usepackage[bitstream-charter]{mathdesign}
\usepackage[capitalize]{cleveref}

\usepackage{xspace}
\usepackage[T1]{fontenc} 
\usepackage{longtable}
\usepackage{graphicx,slashed,multirow,mathtools,sidecap,tikz,bm,enumitem,booktabs,array,makecell}
\usepackage{physics}
\usepackage{multirow}
\usepackage{xspace}
\usepackage{subcaption}
\usepackage{units}
\usepackage{feynmp-auto}
\usepackage{tikz}
\usepackage{tikz-feynman}
\definecolor{cornflowerblue}{rgb}{0.39, 0.58, 0.93}

\usepackage{array}
\newcolumntype{P}[1]{>{\centering\arraybackslash}p{#1}}

\DeclareSymbolFont{usualmathcal}{OMS}{cmsy}{m}{n}
\DeclareSymbolFontAlphabet{\mathcal}{usualmathcal}

\fancypagestyle{SPstyle}{
\fancyhf{}
\lhead{\colorbox{scipostblue}{\bf \color{white} ~SciPost Physics Community Reports}}
\rhead{{\bf \color{scipostdeepblue} ~Submission }}

\fancyfoot[C]{\textbf{\thepage}}
}

\newcommand{\beq}{\begin{eqnarray}}
\newcommand{\eeq}{\end{eqnarray}}

\newcommand{\GeV}{{\ensuremath\,\rm GeV}\xspace}
\newcommand{\TeV}{{\ensuremath\,\rm TeV}\xspace}

\newcommand{\kala}{\kappa_\lambda}

\newcommand{\fb}{\,\, \mathrm{fb}}

\newcommand{\anyH}{\texttt{anyH3}\xspace}
\newcommand{\anyHH}{\texttt{anyHH}\xspace}
\newcommand{\anyBSM}{\texttt{anyBSM}\xspace}

\newcommand{\MBSM}{\ensuremath{M_{\text{BSM}}}}

\begin{document}

\pagestyle{SPstyle}

\begin{flushright}
\texttt{DESY-26-129\\ IFT-UAM/CSIC-26-137\\ FR-PHENO-2026-014}
\end{flushright}

\begin{center}{\Large \textbf{\color{scipostdeepblue}{
Higgs pair production and Higgs potential\\[0.2em]
in arbitrary renormalisable extensions of the Higgs sector
}}}\end{center}

\begin{center}\textbf{
Henning Bahl\textsuperscript{1a}, Johannes Braathen\textsuperscript{2b}, Martin Gabelmann\textsuperscript{3c}, Kateryna Radchenko Serdula\textsuperscript{4d}, Georg Weiglein\textsuperscript{2,5e}
}\end{center}

\begin{center}
{\bf 1}  Institut für Theoretische Physik, Universität Heidelberg, Philosophenweg 16,\\ 61920 Heidelberg, Germany\\[0.2em]
{\bf 2} Deutsches Elektronen-Synchrotron DESY, Notkestr.~85, 22607 Hamburg, Germany\\[0.2em]
{\bf 3} Albert-Ludwigs-Universit{\"a}t Freiburg, Physikalisches Institut, Hermann-Herder-Str.~3, 79104 Freiburg, Germany\\[0.2em]
{\bf 4} Instituto de F\'isica Te\'orica (UAM/CSIC), Universidad Aut\'onoma de Madrid, Cantoblanco, 28049 Madrid, Spain\\[0.2em]
{\bf 5}  II.\  Institut f\"ur  Theoretische  Physik, Universit\"at  Hamburg, Luruper Chaussee 149,\\ 22761 Hamburg, Germany \\
[\baselineskip]
a \href{mailto:bahl@thphys.uni-heidelberg.de}{\small bahl@thphys.uni-heidelberg.de}\,,\\
b \href{mailto:johannes.braathen@desy.de}{\small johannes.braathen@desy.de}\,,\\
c \href{mailto:martin.gabelmann@physik.uni-freiburg.de}{\small martin.gabelmann@physik.uni-freiburg.de}\,,\\
d \href{mailto:kateryna.radchenko@desy.de}{\small kateryna.radchenko@desy.de},\\
e \href{mailto:georg.weiglein@desy.de}{\small georg.weiglein@desy.de}\\
\end{center}

\begin{abstract}
Higgs self-interactions are crucial for probing the up to now largely unknown form of the Higgs potential that is realised in nature.
Higgs boson pair production at the (High-Luminosity) LHC is of prime importance in this context as it provides experimental access to the trilinear Higgs self-interaction
that enters at leading order in this process. The predictions for the trilinear Higgs self-couplings and the shape of the Higgs potential can vastly differ from the predictions of the Standard Model (SM) as a consequence of physics beyond the SM (BSM). If additional particles are relatively light (but in agreement with all existing experimental and theoretical constraints), their effects may not necessarily be accurately described by established effective field theory methods. 
Because of the potentially large modifications compared to the SM predictions the di-Higgs process has a high sensitivity for probing
BSM physics. 
In order to enable the comparison between experiment and theory for a wide range of BSM models, 
automatised predictions for double Higgs production are in order.
In this contribution to the LHC Higgs Working Group Report 5 we present \anyHH, a novel tool for calculating the Higgs boson pair production cross-section in gluon fusion, which is the dominant production process for two Higgs bosons at the (HL-)LHC, for arbitrary renormalisable extensions of the Higgs sector. 
\end{abstract}

\tableofcontents

\section{Introduction}
Higgs boson pair production is among the main targets of the High-Luminosity LHC
(HL-LHC) and future collider 
experiments. This process gives direct access to the  
trilinear self-coupling of the detected Higgs boson that enters the prediction at leading order, and more generally in models with extended Higgs sectors to trilinear scalar couplings involving two of the detected Higgs bosons. In turn, this provides information about the shape of the Higgs potential realised in nature. 
The Higgs self-interactions and the shape of the Higgs potential are directly related to 
the evolution of the early universe and the nature of the electroweak phase transition \cite{Noble:2007kk}. 
While the mass of the  
detected Higgs boson $h$ has now become a precision observable, the remainder of the Higgs potential, in particular the trilinear and quartic self-interactions, is largely unconstrained. 
Large deviations from the predictions of the Standard Model (SM) that are in agreement with all experimental and theoretical constraints~\cite{Bahl:2022jnx} are possible in Beyond-the-SM (BSM) theories.
The
trilinear Higgs coupling modifier $\kappa_\lambda\equiv\lambda_{hhh}/(\lambda_{hhh}^{\text{SM}})^{(0)}$ (where $(\lambda_{hhh}^{\text{SM}})^{(0)}$ is the tree-level prediction in the SM) is
currently bound to be $-0.71 < \kappa_\lambda < 6.1$~\cite{CMS:2026nuu}, while the quartic self-coupling is so far
mostly constrained by theoretical bounds from perturbative unitarity~\cite{DiLuzio:2017tfn,Stylianou:2023tgg}. 
In contrast, the interactions of $h$ with the gauge boson and fermion sector are known
to a remarkable level of accuracy, of $\mathcal{O}(10\%)$, thanks to LHC measurements.  
Despite the fact that the mentioned bounds on $\kappa_\lambda$ are still rather weak, they already probe parameter space of BSM scenarios that has been unconstrained up to now~\cite{Bahl:2022jnx}. 
The Higgs boson pair production process therefore has a high discovery potential for detecting BSM effects. The pattern that BSM contributions to the Higgs self-couplings can be very large while the effects in the gauge and Yukawa sector remain small happens quite naturally in scenarios where deviations from the SM are driven by loop effects
from BSM scalars, as can be shown 
based on power counting arguments, see for instance Ref.~\cite{Bahl:2026aou}. Similar conclusions have been reached in effective field theory analyses~\cite{Durieux:2022hbu}.

State-of-the-art searches 
for Higgs boson pair production can currently be divided into at least three different types:\footnote{We note in this context that several studies have shown that experimental bounds obtained from resonant searches can sometimes not be interpreted reliably, due to significant interference effects that 
up to now have not been taken into account by ATLAS and CMS as part of their signal model, see e.g.\ Refs.~\cite{Heinemeyer:2024hxa,Arco:2025nii,DiHiggsreport}.} (i) non-resonant $hh$-searches, (ii) resonant $hh$ searches and (iii) resonant production in association with at least one BSM scalar. In all of these, scalar couplings of the SM-like Higgs boson to itself 
and possibly to other BSM scalars enter the leading order amplitude. Therefore di-Higgs production cross-sections are very sensitive to loop effects from the BSM sector on scalar couplings. 
As mentioned above, in Ref.~\cite{Bahl:2022jnx} 
it has been shown for the example of the Two-Higgs-Doublet-Model (THDM), that large BSM deviations in $\kala$ are possible for 
parameter regions
that are allowed by 
all existing theoretical and experimental constraints, and that current experimental limits on $\kala$ (derived from limits on the Higgs boson pair production cross-section) are already strong enough to provide additional sensitivity to otherwise unconstrained parameter space. 
Therefore, 
BSM loop contributions to trilinear Higgs couplings are of crucial phenomenological relevance, see also Refs.~\cite{Kanemura:2002vm,Kanemura:2004mg,Aoki:2012jj,Kanemura:2015fra,Kanemura:2015mxa,Hashino:2015nxa,Kanemura:2016lkz,Kanemura:2017wtm,Kanemura:2017gbi,Chiang:2018xpl,Senaha:2018xek,Braathen:2019pxr,Kanemura:2019slf,Braathen:2019zoh,Braathen:2020vwo,Bahl:2022jnx,Bahl:2022gqg,Bahl:2023eau,Bahl:2025wzj,Braathen:2025qxf,Braathen:2025svl,Braathen:2026glt} 
for studies in concrete BSM scenarios including and beyond the THDM.

Due to the large 
variety of BSM models with extended scalar sectors, the public \texttt{Python} program \anyH~\cite{Bahl:2023eau} was developed within the package \anyBSM, which in its original version computes the trilinear self-coupling $\lambda_{sss}$ of a given scalar $s$ at the full one-loop order in any renormalisable theory. More recently, \anyH was extended~\cite{Bahl:2026nsu} to allow full one-loop calculations of any trilinear scalar coupling $\lambda_{ijk}$($\equiv \lambda_{h_ih_jh_k}$, where $h_i$ generically denotes scalars). 

At the level of the Higgs boson pair production cross-section, only few models have been studied so far beyond the SM with the focus set on QCD corrections~\cite{Dawson:1998py,Buchalla:2018yce,Heinrich:2019bkc,Heinrich:2020ckp,Heinrich:2022idm,Heinrich:2023rsd,Baglio:2023euv,Alasfar:2023xpc,Heinrich:2024rtg} (see also e.g.\  Refs.~\cite{Baglio:2020wgt,AH:2022elh,Bagnaschi:2023rbx,Jaskiewicz:2024xkd,Grober:2026ece,H:2026ksi} for recent progress on the calculations of higher-order QCD corrections in the SM, and Refs.~\cite{Davies:2022ram,Muhlleitner:2022ijf,Bi:2023bnq,Davies:2023npk,Zhang:2024rix,Heinrich:2024dnz,Davies:2025wke,Bonetti:2025vfd,Bhattacharya:2025wnr,Davies:2026wbx} for EW corrections in the SM). 
The importance of BSM loop contributions to the Higgs three-point functions has been pointed out mostly in the context of the THDM and singlet extensions of the SM~\cite{Heinemeyer:2024hxa,Feuerstake:2024uxs,Arco:2025pgx,Arco:2025nii,Braathen:2025qxf,Degrassi:2025pqt,Borschensky:2022pfc,Boto:2026bmz%
}.

In this contribution to the LHC Higgs Working Group Report 5, we present a novel tool, \anyHH, also part of the \anyBSM framework and building upon \anyH, that enables  predictions for pair
production cross-sections of two (not necessarily identical) scalars in gluon fusion in a large class of renormalisable models.
As a contribution to the Report 5, whose main purpose is to document the technical challenges and the physics potential of Run 4 and beyond, we focus here on showcasing the phenomenological relevance of \anyHH and its utility for the community. For a detailed technical discussion on the renormalisation procedure, the code and other phenomenological applications we refer the reader to Ref.~\cite{Bahl:2026nsu} and 
further work in progress~\cite{Bahl:2026anyhh}.

The code 
and instructions can be found at the following link:
\begin{center}
\url{https://anybsm.gitlab.io/}
\end{center}

This contribution is organised as follows. In \cref{sec:toymodel} we discuss the general phenomenological features of the total and differential di-Higgs cross-sections using a toy model. \cref{sec:models} demonstrates the flexibility of \anyHH by discussing mass splitting effects and their influence on the total di-Higgs production cross-section in a variety of different renormalisable extensions of the SM with additional scalars charged under $SU(2)_L$. In \cref{sec:thdm} we use the THDM as a concrete model to study in more detail NLO BSM corrections to differential di-Higgs invariant mass distributions in scenarios with and without tree-level alignment. We conclude in \cref{sec:conc}.

\section{General features of Higgs boson pair production}
\label{sec:toymodel}
\begin{figure}[htbp]
\begin{center}
\begin{minipage}{0.3\textwidth}
\begin{tikzpicture}
  \begin{feynman}
    \vertex (g1) at (-2,1) {\(g\)};
    \vertex (g2) at (-2,-1) {\(g\)};
    
    \vertex (a) at (-0.7,0.7);
    \vertex (b) at (-0.7,-0.7);
    \vertex (c) at (0.7,-0.7);
    \vertex (d) at (0.7,0.7);

    \vertex (h1) at (2,1) {\(h_i\)};
    \vertex (h2) at (2,-1) {\(h_j\)};
    
    \diagram* {
      (g1) -- [gluon] (a) -- [fermion] (b) -- [fermion] (c) -- [fermion] (d) -- [fermion] (a),
      (g2) -- [gluon] (b),
      (d) -- [scalar] (h1),
      (c) -- [scalar] (h2),
    };

    \filldraw[olive] (d) circle (2pt) node[below right=0.1pt, olive] {\(c^{t,(0)}_i\)};
    \filldraw[orange] (c) circle (2pt) node[above right=0.1pt, orange] {\(c_j^{t,(0)}\)};

    \node at (-1.1,0) {\(t\)};
    
  \end{feynman}
\end{tikzpicture}
\end{minipage}
\hfill
\begin{minipage}{0.3\textwidth}
\begin{tikzpicture}
  \begin{feynman}
    \vertex (g1) at (-2,1) {\(g\)};
    \vertex (g2) at (-2,-1) {\(g\)};
    \vertex (a) at (-1,0.6);
    \vertex (c) at (-1,-0.6);
    \vertex (b) at (0,0);
    \vertex (h1) at (1.2,0);
    \vertex (h2) at (2.2,0.8) {\(h_i\)};
    \vertex (h3) at (2.2,-0.8) {\(h_j\)};
    
    \diagram* {
      (g1) -- [gluon] (a) -- [fermion] (b) -- [fermion] (c) -- [fermion] (a),
      (g2) -- [gluon] (c),
      (b) -- [scalar, edge label=\(h_k\)] (h1),
      (h1) -- [scalar] (h2),
      (h1) -- [scalar] (h3),
    };
    
    \filldraw[magenta] ($(b)!0!(h1)$) circle (2pt) node[below=3pt, magenta] {\(c_k^{t,(0)}\)};
    \filldraw[cornflowerblue] ($(h1)!0!(h2)$) circle (6pt) node[below=3pt, red] {\(\lambda^{(1)}_{ijk}\)};
    
    \node at (-1.4,0) {\(t\)};
  \end{feynman}
\end{tikzpicture}
\end{minipage}
\hfill
\begin{minipage}{0.3\textwidth}
\begin{tikzpicture}
  \begin{feynman}
    \vertex (g1) at (-2,1) {\(g\)};
    \vertex (g2) at (-2,-1) {\(g\)};
    \vertex (a) at (-1,0.6);
    \vertex (c) at (-1,-0.6);
    \vertex (b) at (0,0);
    \vertex (h1) at (1.2,0);
    \vertex (h2) at (2.2,0.8) {\(h_i\)};
    \vertex (h3) at (2.2,-0.8) {\(h_j\)};
    
    \diagram* {
      (g1) -- [gluon] (a) -- [fermion] (b) -- [fermion] (c) -- [fermion] (a),
      (g2) -- [gluon] (c),
      (b) -- [scalar, edge label=\(h_l\quad h_k\)] (h1),
      (h1) -- [scalar] (h2),
      (h1) -- [scalar] (h3),
    };
    
    \filldraw[purple] ($(b)!0!(h1)$) circle (2pt) node[below=3pt, purple] {\(c_{l}^{t,(0)}\)};
    \filldraw[cornflowerblue] ($(b)!0.5!(h1)$) circle (6pt) node[below=3pt, magenta] {};
    \filldraw[red] ($(h1)!0!(h2)$) circle (2pt) node[below=3pt, red] {\(\lambda^{(0)}_{ijk}\)};
    
    \node at (-1.4,0) {\(t\)};
  \end{feynman}
\end{tikzpicture}
\end{minipage}
\end{center}
\begin{center}
    \parbox{0.95\textwidth}{
        \centering
        \caption{Generic diagrams contributing to Higgs boson pair production via  gluon fusion that are included in \texttt{anyHH}. The first diagram is the box diagram, while the second is the triangle diagram. In the SM, all particles contributing to the second diagram are identical. The third diagram shows the propagator corrections present in the triangle diagram, depicted as a blue blob, which can optionally be included. Generic trilinear Higgs couplings $\lambda_{ijk}$
        can be evaluated at tree-level, $\lambda^{(0)}_{ijk}$, as shown by a red dot in the right diagram, or at one loop, $\lambda^{(1)}_{ijk}$, as shown by a blue blob in the middle diagram. The other couplings involved are the tree-level top Yukawa couplings of the respective scalar particles, depicted here as coloured dots.
        }
         \label{fig:gghhdiagrams}
        }
\end{center}
\end{figure}
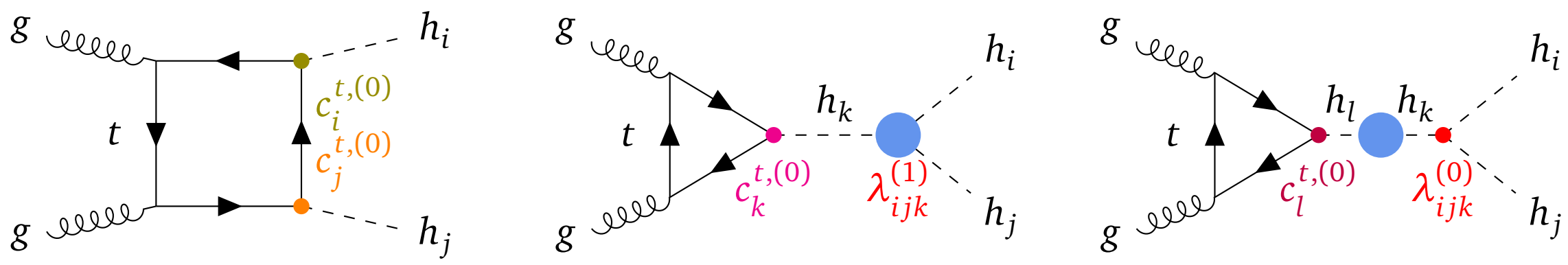

Higgs-pair production via gluon fusion at leading order is mediated by the first two diagrams shown in Fig.~\ref{fig:gghhdiagrams}, corresponding to the box (left) and triangle (middle) topologies. The predictions in \anyHH are computed at leading order in QCD and are applicable to a broad range of renormalisable extensions of the SM. Currently, the implemented leading-order calculation includes only top-quark loops\footnote{Due to this restriction, the calculation is only reliable for SM extensions that do not add extra coloured states.}, which constitute the dominant contributions due to the Yukawa hierarchy. The \texttt{anyBSM} framework also automatically incorporates loop corrections to the trilinear scalar couplings, including their full momentum dependence (blue blob
in \cref{fig:gghhdiagrams} (middle)) and to the $s$-channel propagators (blue blob in \cref{fig:gghhdiagrams} (right)). This is particularly relevant in BSM scenarios with heavy BSM scalar particles, which are known to induce large one-loop corrections to the trilinear scalar couplings in some parameter regions. Furthermore, the final states can be any pair of neutral scalars $h_ih_j$. It is possible to calculate both the total di-Higgs cross-section and the differential distribution as a function of the invariant mass $m_{h_ih_j}$. For the rest of this contribution, we restrict ourselves to the pair production of the detected Higgs boson $h$.

Before showing results in specific BSM theories, we want to briefly showcase the phenomenological behaviour of the Higgs boson pair production predictions in a toy model. This model is SM-like but the relevant quantities for Higgs boson pair production, namely the scalar trilinear coupling $\lambda_{hhh}$ and the top-Yukawa coupling, can be varied freely.

\begin{figure}[h!]
\centering
\includegraphics[width=0.49\textwidth]{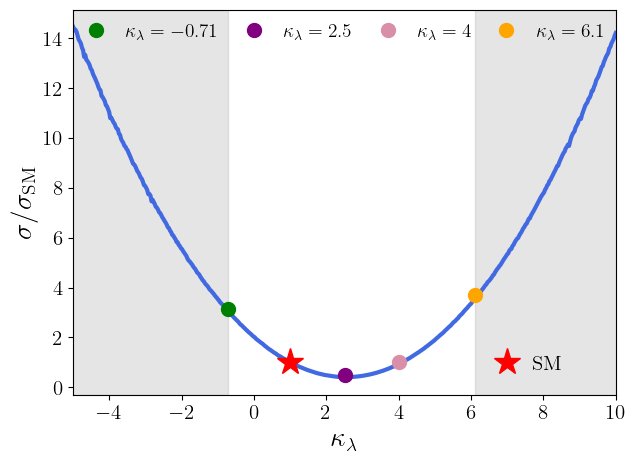}
\includegraphics[width=0.49\textwidth]{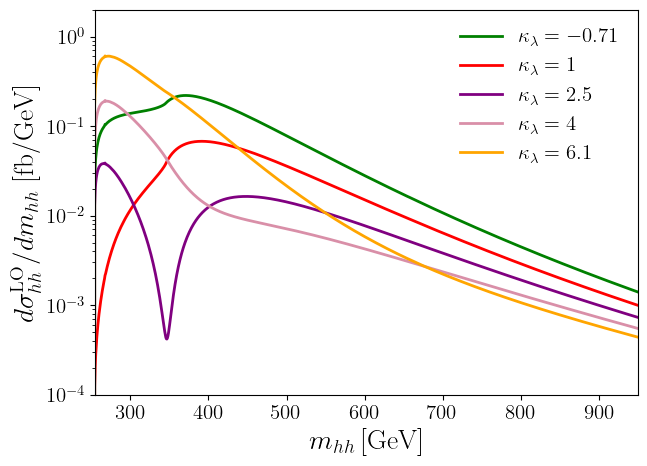}
\includegraphics[width=0.49\textwidth]{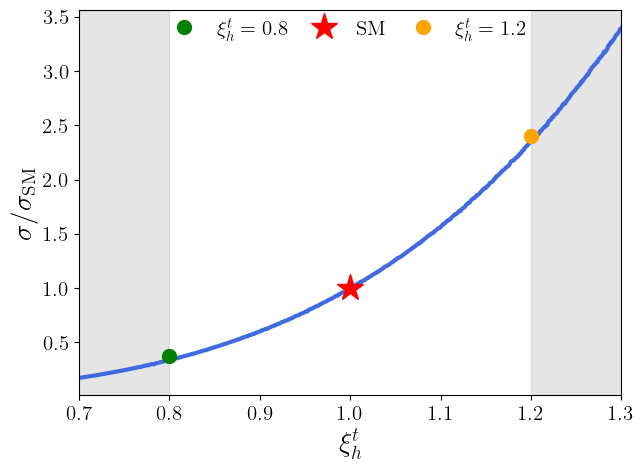}
\includegraphics[width=0.49\textwidth]{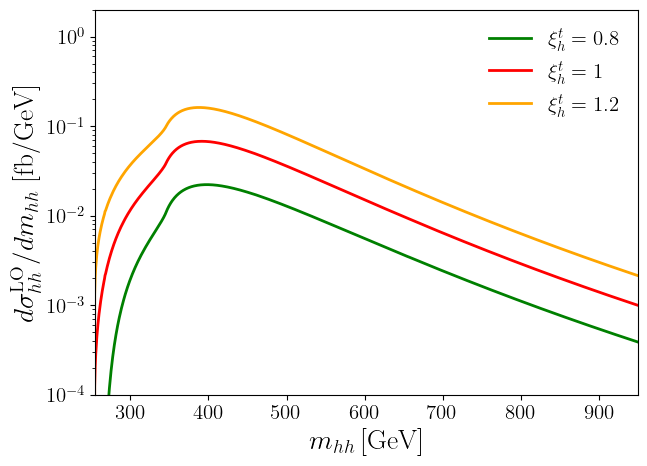}
\caption{Cross-section for non-resonant Higgs boson pair production depending on the trilinear coupling $\kappa_\lambda$ (top) and the top Yukawa coupling modifier $\xi_t^h$ (bottom), for a centre-of-mass energy of $\sqrt{s}=13.6\TeV$. \textit{Left}: total cross-section normalised to the SM LO QCD value calculated with the same \texttt{anyBSM} setup. \textit{Right}: invariant mass distribution of selected benchmark points for a fixed $\kappa_\lambda$ (top) and $\xi_t^h$ (bottom), the value is indicated in the legend.}
\label{fig:toykala}
\end{figure}

In Fig.~\ref{fig:toykala} we show the $\kappa_{\lambda}$ (upper row) and the top Yukawa coupling modifier $\xi_t^h$ (lower row) dependence of the total cross-section (left) and the invariant mass distributions of two SM-like Higgs bosons with 125 \GeV mass in the final state, $m_{hh}$ (right). All quantities are normalised to the SM values: $\kala \equiv \lambda_{hhh}/(\lambda_{hhh}^{\mathrm{SM}})^{(0)}$ and $\xi_h^{t}\equiv (c_h^t)^{(0)}/(c^{t\mathrm{,\,SM}}_{h})^{(0)}$ (using the same notation as in Fig.~\ref{fig:gghhdiagrams} and with the superscript ``$(0)$'' indicating tree-level quantities). The total cross-section is normalised to the SM cross-section at LO QCD calculated within \texttt{anyBSM} (using the same setup as for the toy model in terms of our choice of PDF set and of renormalisation and factorisation scales). 
Throughout this paper we make use of the PDF set \texttt{PDF4LHC15}\footnote{We note that \texttt{PDF4LHC15} is superseded by \texttt{PDF4LHC21}, but this will have a negligible impact on the results shown here. The default in \anyHH is now to use \texttt{PDF4LHC21}. } using \texttt{LHAPDF}~\cite{Buckley:2014ana} and set both, the renormalisation and factorisation scales to one half of the invariant mass of the final state. This choice is not fixed and can be changed in the usage of the code.  
 
The gray regions on the upper left plot show the regions excluded experimentally at 2$\sigma$ by the ATLAS and CMS run 2 combination~\cite{CMS:2026nuu}. We observe a minimum of the total cross-section around the value of $\kala\sim 2.5$, which happens due to the increase in the destructive interference between the triangle and box diagrams. We can see that currently, the maximum possible deviations from the SM cross-section $\sigma_{\mathrm{SM}}$ are roughly a 3-times enhancement for negative $\kala$ and a 4-times enhancement for positive values of $\kala$. We have selected some points to represent their differential cross-section with respect to the invariant mass, $d\sigma^{\mathrm{LO}}_{hh}/d m_{hh}$, we use the symbol $\sigma_{hh}^\mathrm{LO}$ here for the cross-section to make explicit that it is the cross-section in the $hh$-final state and the superscript LO to refer to result at leading order in QCD.\footnote{While applying an NLO-QCD $K$-factor of 2 is justified for the total cross-section, it is less so for the differential invariant mass distribution, see discussion in Ref.~\cite{Heinemeyer:2024hxa}.} These points are the SM case, $\kala = 1$, which is represented with a red star on the left, the extrema of the experimentally allowed interval, $\kala = -0.71$ and $\kala = 6.1$, the value for the maximal destructive interference $\kala = 2.5$, and the value $\kala = 4$, which yields a total cross-section approximately equal to that of the SM.

On the right plot of Fig.~\ref{fig:toykala}, we present the invariant mass distributions corresponding to these points. The SM distribution, shown in red, presents a cancellation of the triangle and box form factors right below the kinematic threshold of $hh$-production, which reduces the contribution to the cross-section from that region. It is later enhanced at around~400 GeV, which corresponds to the di-top production threshold and is also present in single Higgs production via gluon fusion (see e.g. Ref.~\cite{Cepeda:2019klc}). It further decreases for larger di-Higgs invariant masses. If the trilinear coupling is modified to a more negative value as in the case of $\kala = -0.71$ (green curve), the triangle and box diagrams cancel much below the kinematic threshold, which results in an apparent enhancement of the cross-section at threshold. If instead the coupling 
is larger than in the SM, in particular at the minimum where $\kala = 2.5$, the cancellation of the form factors is shifted to higher values of $m_{hh}$, 
which results in a dip at around the di-top threshold, as can be observed in the purple curve. For a value of $\kappa_\lambda = 4$ (pink curve), 
we observe no cancellation near the low-$m_{hh}$ threshold; instead, the differential cross-section is reduced compared to the SM from the di-top threshold.  
The enhancement near the $hh$-production threshold compensates for the suppression around and above the di-top threshold, resulting in the same total cross-section as in the SM. Finally, for the largest allowed value of $\kala=6.1$, the distribution follows a steadily decreasing slope with no significant dips. At threshold the contribution is enhanced over 4 orders of magnitude with respect to the SM and then decreases monotonously.

In the lower row of Fig.~\ref{fig:toykala} we show the dependence of the total (left) and differential (right) cross-section on the SM-like top Yukawa coupling modifier. We again shade the regions of the top Yukawa coupling that are experimentally excluded at 2$\sigma$ C.L.~\cite{CMS:2022dwd} and normalise the total cross-section to the SM prediction, $\sigma_{\mathrm{SM}}$.

In this case we see an increase of about 100\% for the largest possible top Yukawa coupling at $\xi_h^t = 1.2$ and a cross-section reduced by about an order of magnitude for the smallest allowed top Yukawa of $\xi_h^t = 0.8$. We choose these points as representative examples for which to compute the invariant mass distributions, shown in the right plot of Fig.~\ref{fig:toykala}, compared again to the SM distribution in red. We can see that the shape of the invariant mass distribution is not altered by the largest allowed deviations in the top Yukawa coupling, only an overall shift to larger values happens for the 20\% increase of the coupling, corresponding to the increase of about 140\% in the total cross-section, and a similar decrease in the opposite case.

\section{Higgs boson pair production cross-sections for any model}
\label{sec:models}

\begin{figure}
    \centering
    \includegraphics[width=1.0\textwidth]{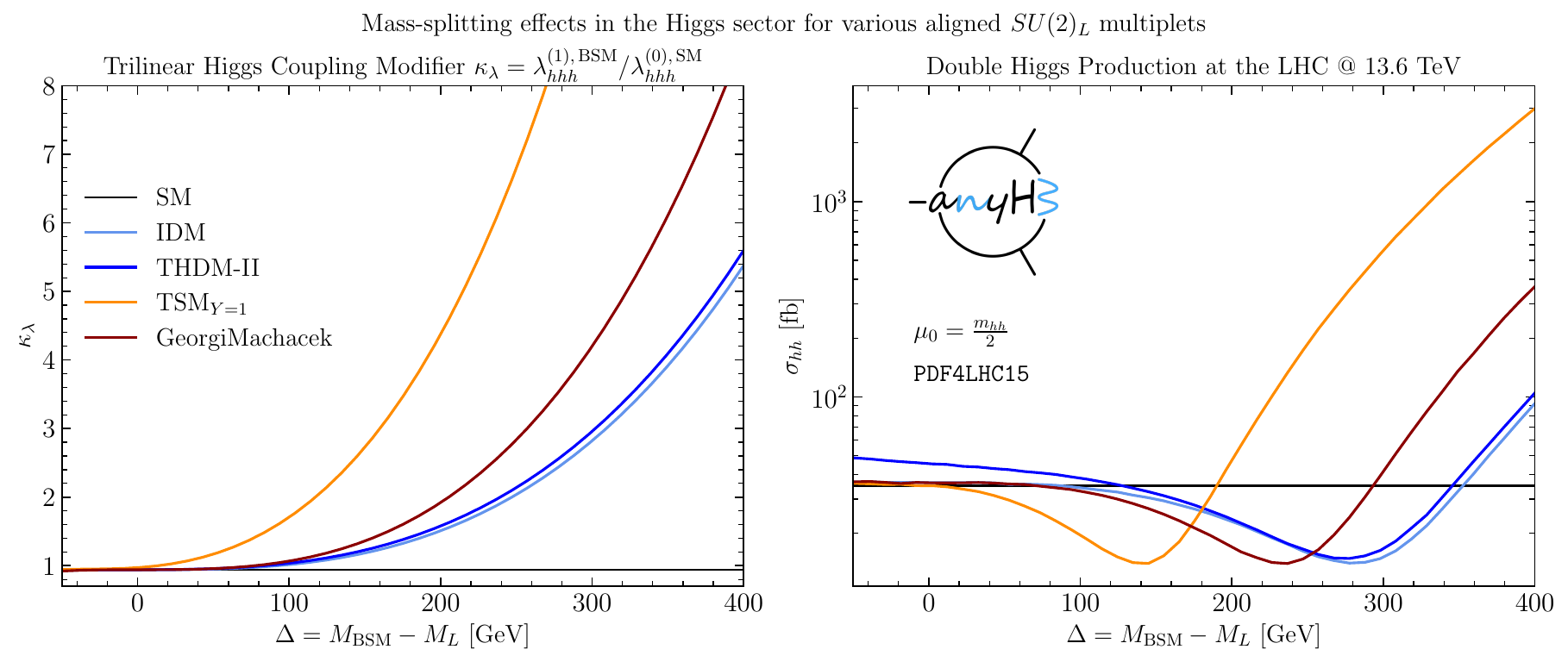}
    \caption{\textit{Left}: Redrawing of Fig. 9 from Ref.~\cite{Bahl:2023eau} with $\Delta = M_{\text{BSM}}-M_L$ where the light (heavy) BSM mass scale $M_L=400\,$GeV ($M_{\text{BSM}}$) along with other input parameters is defined in Ref.~\cite{Bahl:2023eau}. \textit{Right}: The total production cross-section of two SM-like Higgs bosons in gluon fusion using the BSM scenarios shown on the left (same colour scheme).
    } 
    \label{fig:dihiggsModels}
\end{figure}

In Ref.~\cite{Bahl:2023eau} it was shown that BSM contributions to $\lambda_{hhh}$ do not need to decouple when individual BSM states feature large mass splittings. In the following we shortly summarise the origin of these large mass splitting effects, which was already discussed in Ref.~\cite{Bahl:2023eau}, and investigate their effect onto the di-Higgs production cross-section in various scalar extensions of the SM. We use the same benchmark models as in Sec. 6.3 of Ref.~\cite{Bahl:2023eau}: the Inert-Doublet-Model (IDM), the Two-Higgs-Doublet-Model of type II (THDM-II), an inert Triplet Model with hypercharge (TSM$_{Y=1}$) and the inert Georgi-Machacek model (GM). In each model, the lightest BSM state charged under $SU(2)_L$ has a mass $M_L=400\GeV$. For completeness we repeat here the model-specific choices of Ref.~\cite{Bahl:2023eau},
\begin{itemize}
    \item IDM: $M_H=\mu_2=M_L$;
    \item THDM-II: $M=M_H=M_L$, $\tan\beta=2$ and $\cos(\beta-\alpha)=0$;
    \item TSM$_{Y=1}$: $m_{D^{++}}=M_L$, $\lambda_4=2.5$;
    \item GM: $M_{h_2}=M_{\eta}=M_L$, $M_2=v_T=\sin\theta_H=0$;
\end{itemize}
For detailed definitions of all models and model parameters we refer the reader to the appendix of Ref.~\cite{Bahl:2023eau}.
All remaining BSM mass parameters are set to a common value $\MBSM$. In every scenario, the Higgs sector features alignment, meaning that all tree-level couplings of the detected Higgs boson are exactly as in the SM. We note that this also implies $\kappa_\lambda=1$ at tree level.

The decoupling theorem applies when couplings do not grow with the BSM mass scale $\MBSM$. However, this assumption fails if a heavy scalar $X$ gets a significant fraction of its mass from the electroweak vacuum expectation value (VEV) $v$, because the portal coupling $\lambda_{XH}$ (which couples two $X$ and two doublets) then grows with the squared-mass splitting $\lambda_{XH}=2(\MBSM^2-M_L^2)/v^2$.
As a result, loop corrections to $\kappa_\lambda$ (which depend on $\lambda_{HX}$) rise steeply with $\Delta=\MBSM-M_L$ in all four considered models and drive $\kappa_\lambda$ to large values, see the left panel of \cref{fig:dihiggsModels}, which reproduces Fig.~9 in  Ref.~\cite{Bahl:2023eau}.\footnote{We want to stress that perturbative unitarity is still maintained in the shown range of $\Delta$.}
However, experimental limits on $\kappa_\lambda$ are extracted from non-resonant di-Higgs searches. 

The code \anyHH provides the full cross-section prediction for di-Higgs production in gluon fusion once given \texttt{UFO}~\cite{Darme:2023jdn} model files for the considered renormalisable theory, 
including corrections to trilinear scalar coupling computed by \anyH. 
In general, the trilinear self-coupling of the detected Higgs boson, $\lambda_{hhh}$, is not the only relevant trilinear coupling that enters this process: $s$-channel diagrams feature couplings of the form $\lambda_{hhS}$, where $S$ should be summed over all scalars in the model.
Moreover, the code is also able to obtain precise predictions for arbitrary two-scalar final states $gg\to h_jh_j$ (see \cref{fig:toykala}). Therefore, \anyH has been extended to obtain full one-loop predictions to arbitrary trilinear couplings $\lambda_{ijk}$ and to allow arbitrary external squared momenta. It also features a semi-automatic and flexible renormalisation procedure --- in the same spirit as the first version of the program. The loop-corrected trilinear couplings are then used in the calculation of $\sigma(gg\to h_i h_j)$, either as effective couplings (neglecting their momentum dependence) or as momentum-dependent form factors.

We now discuss the result for the total di-Higgs production cross-sections obtained in the aforementioned BSM scenarios, which are shown in the right panel of \cref{fig:dihiggsModels}. 
For $\Delta\lesssim 0\GeV$, the inert models (i.e.\ the IDM, TSM$_{Y=1}$ and the inert GM model) reproduce the SM value of about $37.5\fb$ while the THDM-II predicts cross-sections slightly above the SM.
As $\kappa_\lambda$ grows, destructive interference between the triangle and box diagrams lowers $\sigma_{hh}$ to a minimum of about $14\fb$, at $\kappa_\lambda\approx 2.3-2.5$.
For larger values of $\Delta$, and thus of $\kala$, the squared triangle term (c.f.\ the middle diagram of \cref{fig:gghhdiagrams}) dominates: the cross-section grows roughly like $\kappa_\lambda^2$, passes the SM value again and reaches maximum values of more than one order of magnitude above the SM prediction.

The \textit{inert} models all have a very similar behaviour for $\Delta\lesssim 0\GeV$: in these models, alignment is enforced by symmetry rather than imposed by hand, and it holds to all orders.
The new scalars carry no VEV, do not mix with the SM-like Higgs boson $h$, couple to $h$ only in pairs, and do not couple to quarks.
Consequently, they change $\sigma(gg\to hh)$ at the next-to-leading order mainly through changes in $\kappa_\lambda$ (as well as through the momentum-dependent Higgs propagator corrections, which are also automatically included in \anyHH).
There is no $s$-channel exchange of a heavy scalar and more importantly there are also no double-box or two-loop triangle diagrams with BSM states in the loop.
Therefore, the presented results already constitute a full NLO-BSM prediction while the NLO-QCD corrections factorise and the NLO-electroweak corrections are the same as in the SM.
Thus, $\kappa_\lambda$ can be considered to be a \textit{good} pseudo-observable in the inert scenarios, because the BSM contributions to Higgs boson pair production only arise via modifications in the trilinear Higgs self-coupling.

In the THDM-II (dark-blue line), the $\mathbb{Z}_2$ symmetry is softly broken and both doublets obtain VEVs during electroweak symmetry breaking. The alignment limit $\cos(\beta-\alpha)=0$ is imposed \textit{artificially} at the tree level and no symmetry protects it. As a result, the $\lambda_{hhH}$ coupling vanishes at tree level but is in general non-zero at the loop-level, i.e.\ the resonant $H$ $s$-channel contribution becomes loop-generated. With the extension of \anyH, the loop-induced $\lambda_{hhH}$ coupling is automatically taken into account in the prediction for di-Higgs production. Moreover, the heavy $\mathcal{CP}$-even scalar couples to top quarks even in the alignment limit. Therefore, all types of diagrams in \cref{fig:toykala} receive BSM contributions. This effect is visible in the right panel of \cref{fig:dihiggsModels}: for small $\Delta$, the THDM-II prediction lies above the SM one, while the results in the IDM (and all other inert models) stay close to the SM value although the THDM-II and IDM scenarios have almost the same $\kappa_\lambda$. 
For the THDM benchmark scenario discussed in this section, the resonant effects are still relatively small and become negligible for larger $\Delta$. Therefore, $\kappa_\lambda$ can still be considered as a good approximation in this scenario. However, this conclusion is highly dependent on the considered model and benchmark scenario, see \cref{sec:thdm}. With the program \anyHH it is now possible to evaluate whether $\kappa_\lambda$ is a good pseudo-observable for a given scenario or whether resonant contributions are important as well.

\section{Example scenarios for Higgs boson pair production in the THDM}
\label{sec:thdm}
In this section, we present results for scenarios in a specific model, choosing for this purpose the widely studied $\mathcal{CP}$-conserving THDM~\cite{Lee:1973iz,Branco:2011iw}. The considered scenarios are defined for the THDM of Yukawa-type I, and have all been verified against searches for BSM scalars and measurements of properties of the detected Higgs boson, using \texttt{HiggsTools}~\cite{Bahl:2022igd,Bahl:2026yal}, and found to be allowed. The renormalisation scheme that we adopt is closely inspired by the KOSY scheme of Ref.~\cite{Kanemura:2004mg}, meaning that all parameters but the BSM mass parameter $M$ are renormalised on-shell. The only difference with the original KOSY scheme is that we additionally include the running of $M$ from $Q=M$ to $Q=m_t$, where the calculation of the trilinear scalar couplings is performed. We note that by default the width of the scalars in the $s$-channel propagator contributions are computed by \texttt{anyHH} at leading order using the optical theorem, but these can also be given as external inputs. In this section, we set the centre-of-mass energy for the calculations in \anyHH to $\sqrt{s}=14\TeV$.

\begin{figure}
    \centering
    \includegraphics[width=\textwidth]{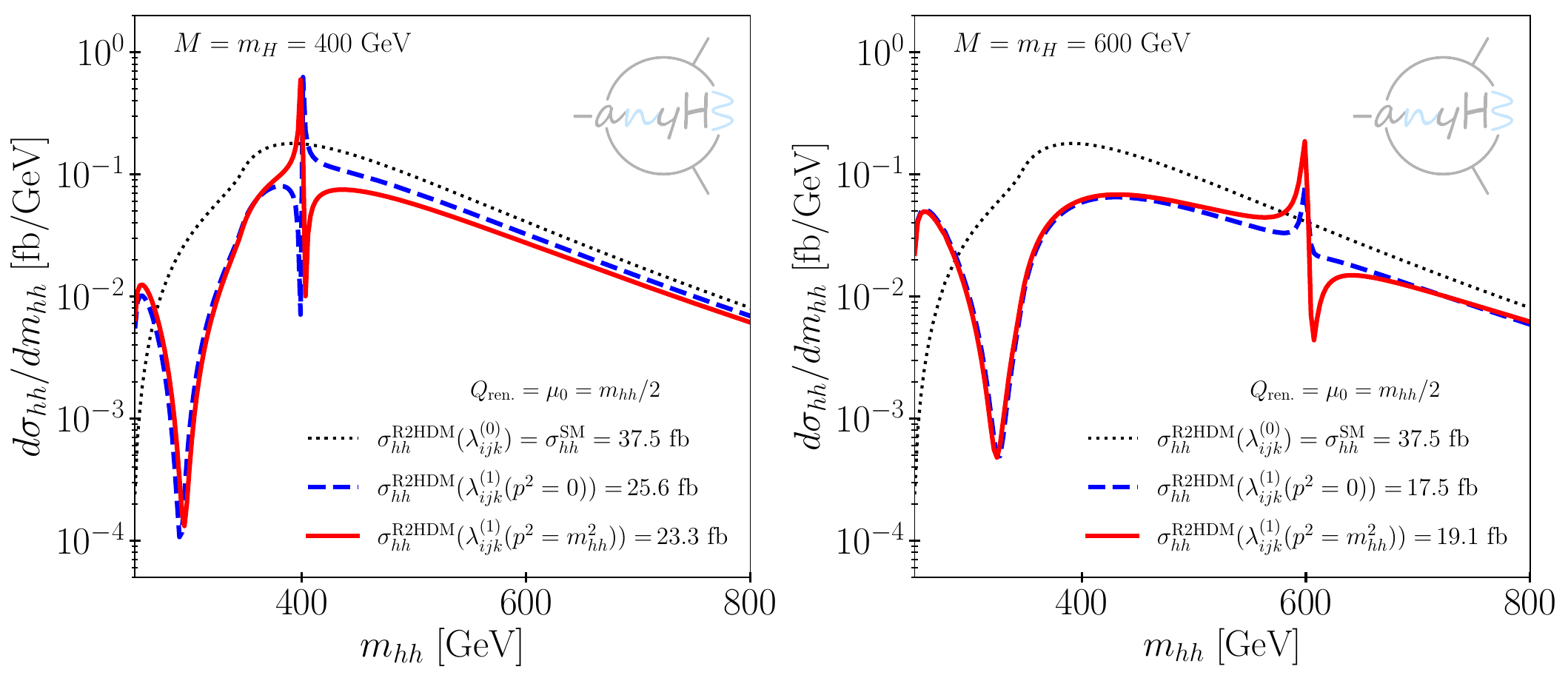}
    \caption{Differential $m_{hh}$ distributions computed with \texttt{anyHH} for THDM scenarios with resonances at different masses. \textit{Left}: $M=m_H=400\text{ GeV}$, $m_A=m_{H^\pm}=600\text{ GeV}$, $\tan\beta=2$, $c_{\beta-\alpha}=0$. \textit{Right}: $M=m_H=600\text{ GeV}$, $m_A=800\text{ GeV}$, $m_{H^\pm}=820\text{ GeV}$, $\tan\beta=2$, $c_{\beta-\alpha}=0$. The black dotted lines are obtained using tree-level values of the trilinear scalar couplings, while the blue dashed lines are obtained with effective one-loop corrected values of the couplings (neglecting momentum dependence). The red solid lines include radiative corrections to the trilinear scalar couplings with full momentum dependence, as well as propagator corrections. Results are shown in this figure, and in the rest of this section for $\sqrt{s}=14\TeV$ (motivated by the target energy at HL-LHC). We have verified that the distributions would be very similar for $\sqrt{s}=13.6\TeV$, with only an overall reduction of about $5-10\%$.}
    \label{fig:2HDM_mhh_varres}
\end{figure}

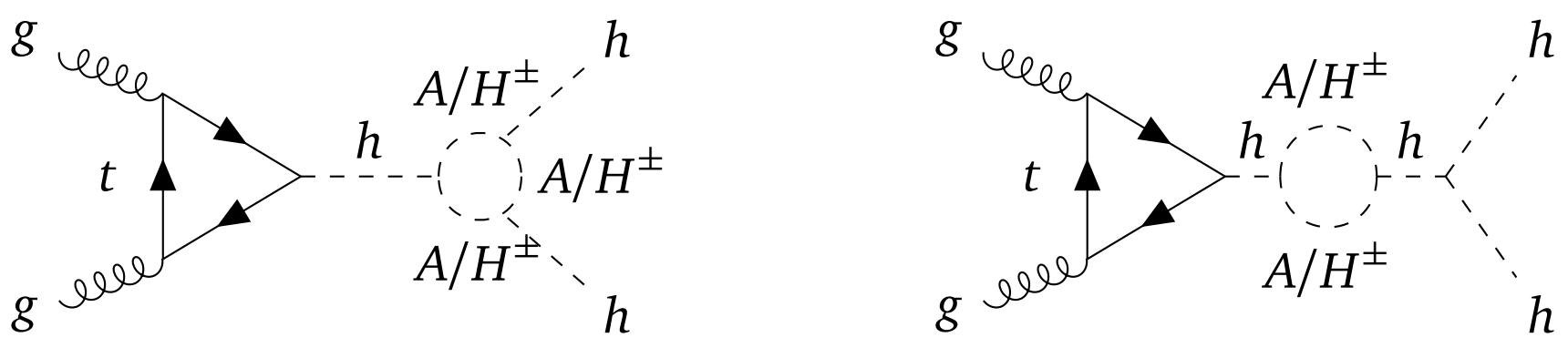
\begin{figure}[htbp]
\begin{center}
\begin{minipage}{0.3\textwidth}
\begin{tikzpicture}
  \begin{feynman}
    \vertex (g1) at (-2,1) {\(g\)};
    \vertex (g2) at (-2,-1) {\(g\)};
    \vertex (a) at (-1,0.6);
    \vertex (c) at (-1,-0.6);
    \vertex (b) at (0,0);
    \vertex (h1) at (1,0);
    \vertex (s) at (1.6,0);
    \vertex (s1) at (1.5,0.3);
    \vertex (s2) at (1.5,-0.3);
    \vertex (h2) at (2.3,1) {\(h\)};
    \vertex (h3) at (2.3,-1) {\(h\)};
    
    \diagram* {
      (g1) -- [gluon] (a) -- [fermion] (b) -- [fermion] (c) -- [fermion] (a),
      (g2) -- [gluon] (c),
      (b) -- [scalar, edge label=\(h\)] (h1),
      (h1) -- [scalar, half left, looseness=1.8, edge label=\(A/H^{\pm}\)] (s) -- [scalar, half left, looseness=1.8, edge label=\(A/H^{\pm}\)] (h1),
      (s1) -- [scalar] (h2),
      (s2) -- [scalar] (h3),
    };
    
    \node at (2.2,0) {\(A/H^{\pm}\)};
    \node at (-1.4,0) {\(t\)};
  \end{feynman}
\end{tikzpicture}
\end{minipage}
\hspace{2cm}
\begin{minipage}{0.3\textwidth}
\begin{tikzpicture}
  \begin{feynman}
    \vertex (g1) at (-2,1) {\(g\)};
    \vertex (g2) at (-2,-1) {\(g\)};
    \vertex (a) at (-1,0.6);
    \vertex (c) at (-1,-0.6);
    \vertex (b) at (0,0);
    \vertex (h1) at (0.4,0);
    \vertex (s) at (1.1,0);
    \vertex (s0) at (1.6,0);
    \vertex (h2) at (2.3,1) {\(h\)};
    \vertex (h3) at (2.3,-1) {\(h\)};
    
    \diagram* {
      (g1) -- [gluon] (a) -- [fermion] (b) -- [fermion] (c) -- [fermion] (a),
      (g2) -- [gluon] (c),
      (b) -- [scalar, edge label=\(h\)] (h1),
      (h1) -- [scalar, half left, looseness=1.8, edge label=\(A/H^{\pm}\)] (s) -- [scalar, half left, looseness=1.8, edge label=\(A/H^{\pm}\)] (h1),
      (s) -- [scalar, edge label=\(h\)] (s0),
      (s0) -- [scalar] (h2),
      (s0) -- [scalar] (h3),
    };
    \node at (-1.4,0) {\(t\)};
  \end{feynman}
\end{tikzpicture}
\end{minipage}
\end{center}
\caption{Example diagrams with possible large contributions
to BSM Higgs boson pair production in the alignment limit of the THDM. The trilinear-vertex and the propagator corrections can contain either the $\mathcal{CP}$-odd BSM scalar $A$ or the charged scalar $H^\pm$. In the predictions of \anyHH, all possible Feynman diagrams are included. }
\label{fig:feyndiags_THDM}
\end{figure}

A first direction we investigate is the impact of different masses of the BSM scalars on the total cross-sections and di-Higgs invariant mass distributions. Our results for the $m_{hh}$ distributions computed with \texttt{anyHH} are displayed in \cref{fig:2HDM_mhh_varres}: the left panel for a scenario with $M=m_H=400\text{ GeV}$, $m_A=m_{H^\pm}=600\text{ GeV}$ and the right one with $M=m_H=600\text{ GeV}$, $m_A=800\text{ GeV}$ and $m_{H^\pm}=820\text{ GeV}$. Both scenarios are defined in the alignment limit, i.e.\ $c_{\beta-\alpha}(\equiv\cos(\beta-\alpha))=0$, and with $\tan\beta=2$. The different line styles indicate the order at which the relevant trilinear scalar couplings (namely, $\lambda_{hhh}$ and $\lambda_{hhH}$) are computed: black dotted lines correspond to tree-level trilinear couplings; blue dashed lines to one-loop corrected trilinear couplings evaluated at vanishing external momenta; and red solid lines are the predictions obtained when including the full momentum dependence of the trilinear couplings --- evaluated with the two external legs on shell, $p^2=m_h^2$, and the one leg connected to an internal propagator with $p^2=m_{hh}^2$ --- and in the $s$-channel propagator corrections. We note that in this figure, the black dotted curves are identical to the SM prediction, due to the alignment limit that implies tree-level values of $\kappa_\lambda^{(0)}=1$ and $\lambda_{hhH}^{(0)}=0$. Once radiative corrections are included in the di-Higgs predictions (blue and red curves), the $m_{hh}$ distributions are significantly modified, as can be observed for both scenarios considered in \cref{fig:2HDM_mhh_varres}. Some example Feynman diagrams of the most relevant types of contributions are presented in \cref{fig:feyndiags_THDM} --- we recall that the loop contributions from the $\mathcal{CP}$-even BSM scalar $H$ are small in these scenarios, because we fix $M=m_H$, which leads to a vanishing $\lambda_{hHH}$ coupling. 

Two main effects can be distinguished on the distributions: first, near the di-Higgs threshold (around $m_{hh}\gtrsim 2m_h$) the change in $\kappa_\lambda$ strongly modifies the interference between the SM-like triangle diagram (middle diagram of \cref{fig:gghhdiagrams} with $\lambda_{hhh}$) and the box diagram (left diagram of \cref{fig:gghhdiagrams}). The accidental cancellation between the two contributions, which occurs in the SM case near the di-Higgs threshold, is lifted, and instead the location of the maximal destructive interference between triangle and box diagrams is shifted to higher values of $m_{hh}$, as also discussed in \cref{sec:toymodel}. This behaviour is more pronounced for the point with the heavier resonance, which can be understood by the larger one-loop value of $\kala^{(1)}$ for this point ($\sim 2.1$ compared to $\sim 1.6$ for the point in the left panel, see \cref{tab:THDM_varres_tri}). The other clear consequence of including loop-corrected trilinear couplings is the appearance of a resonance peak around $m_{hh}\simeq m_H$; this peak is absent when using tree-level couplings because $\lambda_{hhH}^{(0)}=0$, due once more to the alignment condition (we discuss below scenarios away from the alignment limit). Interestingly, we observe in the left panel that the inclusion of the momentum dependence of the trilinear couplings can have a significant impact on the structure of the resonance: indeed, without momentum (blue curve) the resonance has a dip-peak structure, while once momentum effects are taken into account (red curve), the resonance structure becomes peak-dip. This can be understood by comparing the values of the $\lambda_{hhH}$ entering the calculations of the two predictions, c.f.\ \cref{tab:THDM_varres_tri}: for vanishing external momenta, the value of $\lambda_{hhH}$ is slightly negative, while when evaluated for momenta equal to $(m_h^2,m_h^2,m_H^2)$ --- corresponding to the two external $h$ legs of the $gg\to hh$ diagram being on-shell and the internal $H$ propagator at $m_{hh}^2=m_H^2$, i.e.\ at the location of the resonance --- its value is slightly positive. This shows that for benchmark points in the alignment limit (or sufficiently close to it), so that the tree-level value of $\lambda_{hhH}$ is small, the impact of momentum effects can be very important. Such a behaviour is, however, not found for the benchmark point in the right panel of \cref{fig:2HDM_mhh_varres}, for which $\lambda_{hhH}$ is negative for both vanishing and non-zero external momenta (see \cref{tab:THDM_varres_tri}). Lastly, we can consider the total cross-section predictions for these example points --- our results are quoted in the legends of the plots in \cref{fig:2HDM_mhh_varres}. For both scenarios, the inclusion of radiative corrections in the trilinear scalar couplings leads to a noticeable decrease in the Higgs boson pair production cross-sections compared to the SM result. This is mainly due to the modification of the interference pattern between the triangle and box contributions from the increase in $\kala$. This agrees with the dependence discussed in \cref{sec:toymodel} and with the discussion of \cref{fig:dihiggsModels} in \cref{sec:models} (see also Ref.~\cite{LHCHiggsCrossSectionWorkingGroup:2016ypw}). The more moderate differences between blue and red curves stem from the structure and precise shapes of the resonances, which also affect the total cross-sections.   

\begin{table}[ht]
\centering
\begin{tabular}{|l|ccccc|}
\hline
 & $\kala^{(0)}$ & $\kala^{(1)}$ & $\lambda_{hhH}^{(0)}$ & $\lambda_{hhH}^{(1)}(p^2=0)$ & $\lambda_{hhH}^{(1)}(m_h^2,m_h^2,m_H^2)$ \\
 \hline
 $M=m_H=400\text{ GeV}$, & 1 & 1.58 & 0 & 5.78 GeV& $-6.81$ GeV \\
 $m_A=m_{H^\pm}=600\text{ GeV}$ & & & & & \\
 \hline
 $M=m_H=600\text{ GeV}$, & & & & & \\
 $m_A=800\text{ GeV}$, & 1 & 2.09 & 0 & $-8.06$ GeV & $-19.60$ GeV\\
 $m_{H^\pm}=820\text{ GeV}$ & & & & & \\
\hline
\end{tabular}
\caption{Values of the trilinear scalar couplings, at tree level and one loop, for the benchmark points used in \cref{fig:2HDM_mhh_varres}.}
\label{tab:THDM_varres_tri}
\end{table}

\begin{figure}
    \centering
    \includegraphics[width=\textwidth]{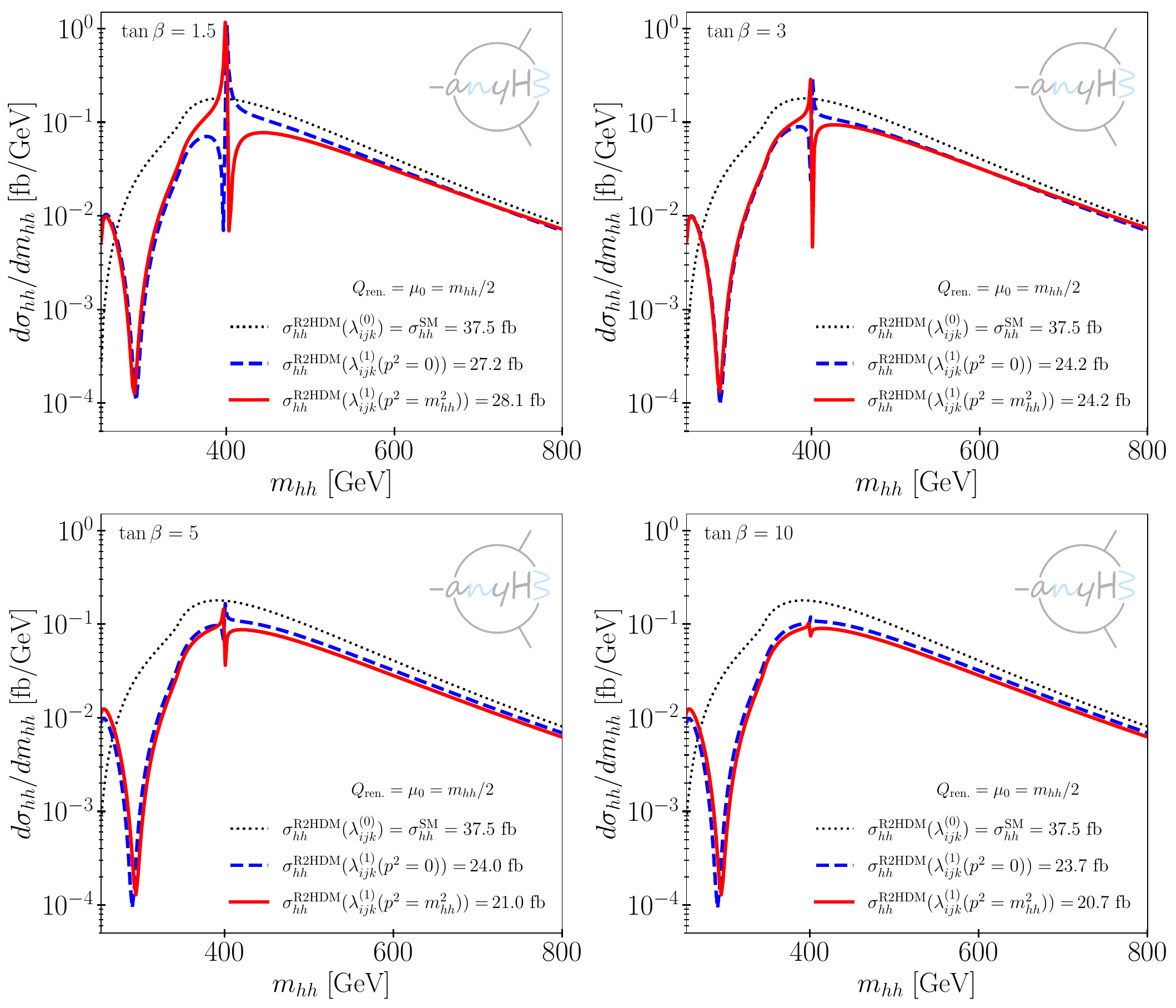}
    \caption{Differential $m_{hh}$ distributions computed with \texttt{anyHH} (for $\sqrt{s}=14\TeV$) for THDM scenarios defined by $M=m_H=400\text{ GeV}$, $m_A=m_{H^\pm}=600\text{ GeV}$, $s_{\beta-\alpha}=1$ and different values of $\tan\beta$. \textit{Upper left}: $\tan\beta=1.5$; \textit{upper right}: $\tan\beta=3$; \textit{lower left:} $\tan\beta=5$; \textit{lower right}: $\tan\beta=10$. } 
    \label{fig:2HDM_mhh_vartanb}
\end{figure}

Next, in \cref{fig:2HDM_mhh_vartanb}, we investigate the dependence of the Higgs pair production predictions on $\tan\beta$. We start from a benchmark point similar to that in the left panel of \cref{fig:2HDM_mhh_varres}, i.e.\ $M=m_H=400 \text{ GeV}$, $m_A=m_{H^\pm}=600\text{ GeV}$, and $c_{\beta-\alpha}=0$, but we present results for different values of $\tan\beta\in\{1.5,3,5,10 \}$. The colour coding of the various curves is the same as in \cref{fig:2HDM_mhh_varres}. The behaviour of the blue and red curves (using loop-corrected trilinear scalar couplings) near the di-Higgs threshold is unchanged between the different values of $\tan\beta$. This is explained by the fact that $\kala^{(1)}$ is essentially independent from $\tan\beta$ in the alignment limit of the THDM of type I, as Yukawa and gauge couplings as well as the large scalar couplings are all independent of $\tan\beta$. On the other hand, the loop-corrected $\lambda_{hhH}$ coupling does exhibit a dependence on $\tan\beta$, and larger values of $\tan\beta$ lead to smaller absolute values of the coupling (as in the case discussed above, for the points considered here, the sign of the coupling depends on whether non-vanishing external momenta are considered or not). Due to this reduction of $|\lambda_{hhH}|$, the resonance peak becomes more difficult to distinguish for larger $\tan\beta$. If experimental effects (like detector uncertainties or binning) are taken into account, it seems unlikely that a resonance could be discriminated from the continuum contribution for the case of $\tan\beta=10$ (and likely also for $\tan\beta=5$, depending on what magnitude of effects are considered). In turn, the reduced size of the resonance peaks also leads to smaller total cross-sections for increasing $\tan\beta$. 

\begin{figure}
    \centering    \includegraphics[width=\textwidth]{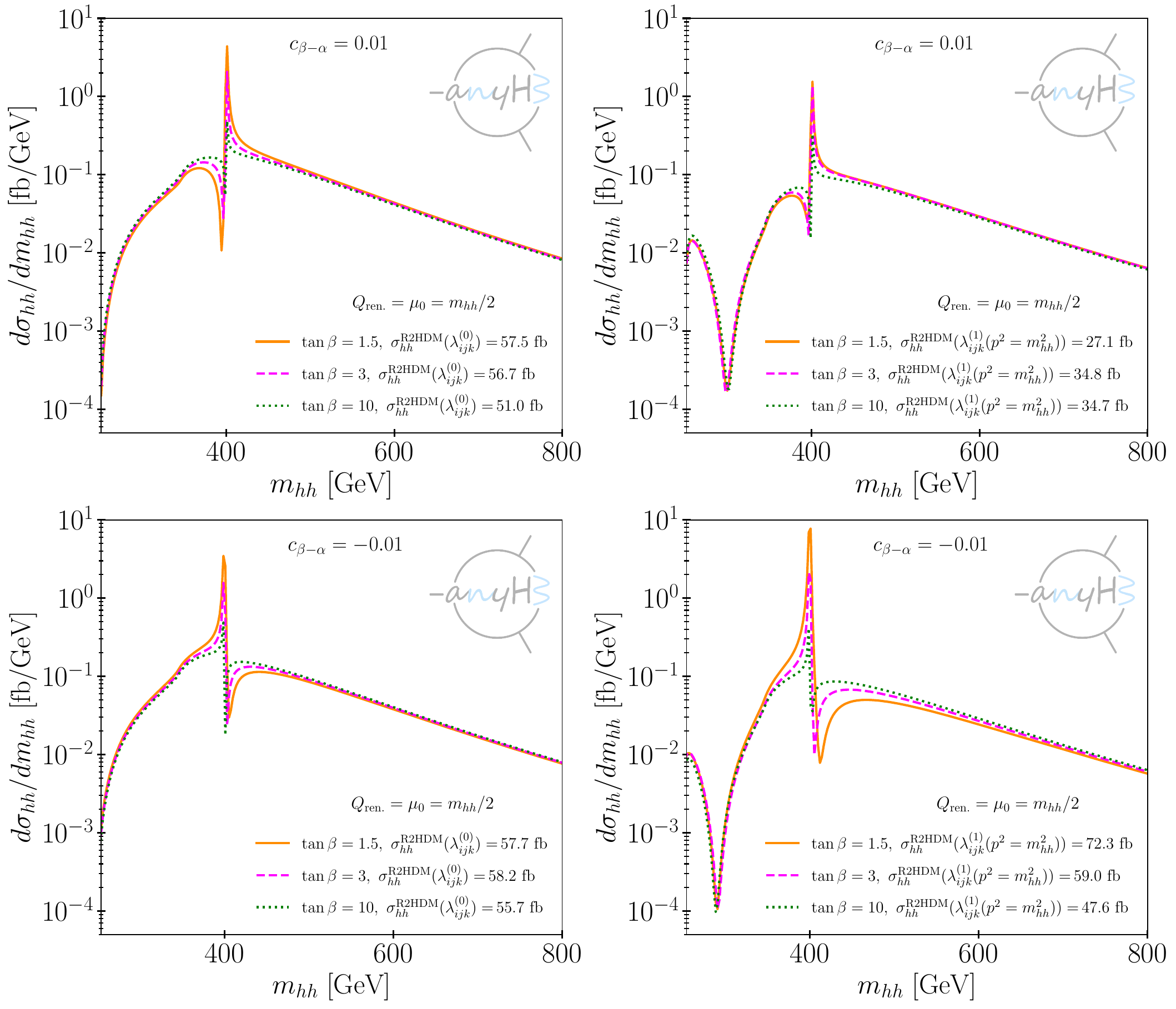}
    \caption{Differential $m_{hh}$ distributions computed with \texttt{anyHH} (for $\sqrt{s}=14\TeV$) for THDM scenarios defined by $M=m_H=400\text{ GeV}$, $m_A=m_{H^\pm}=600\text{ GeV}$, and different values of $c_{\beta-\alpha}$ and $\tan\beta$. The upper (lower) row plots correspond to $c_{\beta-\alpha}=0.01$ ($c_{\beta-\alpha}=-0.01$). The left (right) column plot are results calculated with tree-level (one-loop corrected and momentum-dependent) trilinear scalar couplings. The different line style corresponds to different values of $\tan\beta$: orange solid for $\tan\beta=1.5$, magenta dashed for $\tan\beta=3$ and green dotted for $\tan\beta=10$. }
    \label{fig:THDM_wmixing}
\end{figure}

Finally, we explore the impact of deviations from the alignment limit. We consider once more scenarios with $M=m_H=400\text{ GeV}$ and $m_A=m_{H^\pm}=600\text{ GeV}$, and vary $c_{\beta-\alpha}=\pm 0.01$ and $\tan\beta\in\{1.5,3,10\}$. Our results are displayed in \cref{fig:THDM_wmixing}: the upper (lower) row plots are for $c_{\beta-\alpha}=0.01$ ($-0.01$). The left plots are obtained using tree-level trilinear couplings, while the right plots employ loop-corrected and momentum-dependent predictions. The colours correspond to different values of $\tan\beta$, as indicated in the plot legends. The behaviour of the various curves near the di-Higgs threshold is very similar to the previous examples shown in the alignment limit, and the increase of $\kala$ due to radiative corrections from the BSM scalars causes a shift of the location of the maximal interference between triangle and box diagrams. Comparing positive and negative values of $c_{\beta-\alpha}$, we find that the structure of the resonance peak is changed from dip-peak (for $c_{\beta-\alpha}>0$) to peak-dip (for $c_{\beta-\alpha}<0$). This can be understood by the linear dependence of the tree-level expression of $\lambda_{hhH}^{(0)}$ on $c_{\beta-\alpha}$ (see e.g.\ Ref.~\cite{Kanemura:2004mg}). Once radiative corrections are included, we find overall that the size of the resonance peak is reduced for $c_{\beta-\alpha}=0.01$ while it is increased for $c_{\beta-\alpha}=-0.01$. 
The change in the total cross-section depends on both $c_{\beta-\alpha}$ and $\tan\beta$ with reductions of up to $\sim 50\%$ (for $c_{\beta-\alpha}=0.01$ and $\tan\beta=1.5$) and increases of up to $\sim 25\%$ (for $c_{\beta-\alpha}=-0.01$ and $\tan\beta=1.5$). 

The examples presented in this section illustrate the importance of including higher-order corrections to the trilinear scalar couplings relevant for the Higgs boson pair production process, as can now be done for a broad range of BSM models with extended scalar sectors using the public code \texttt{anyHH}. 

\section{Conclusions}
\label{sec:conc}
Higgs pair production is a key process to access the shape of the Higgs potential that is realised in nature and to probe BSM physics. 
At the same time, a 
wide range of BSM scenarios exists that significantly modify the Higgs sector. In this contribution to the LHCHWG Report 5, we have introduced \anyHH, a new public \texttt{Python} tool built upon \anyH~within the \anyBSM~framework, that provides precise predictions for the production of final states with two Higgs bosons in gluon fusion. Predictions are provided for the total cross-section as well as differentially in the invariant final-state mass.

As a first step in this report we have discussed,
using a toy model 
with freely floating coupling modifiers for the trilinear self-coupling and top Yukawa coupling, the general features of di-Higgs production and its sensitivity to modifications of the two couplings entering the leading-order prediction.

Next, we demonstrated the flexibility of the program by considering a variety of non-trivial BSM theories with extended scalar sectors and we showed the strong dependence of the di-Higgs cross-section on mass splitting effects entering dominantly via loop-corrections to the trilinear Higgs coupling.

Furthermore, we used \anyHH to investigate the impact on the total di-Higgs production cross-section and its differential $m_{hh}$ distribution of the most important model parameters of the
THDM type I: namely the
mass of the $\mathcal{CP}$-even BSM scalar $H$, $\tan\beta$, and the parameter indicating the deviation from the alignment limit, $\cos({\beta-\alpha})$. We found that BSM contributions to $\kala$ have a strong impact on both the $m_{hh}$ distribution and the total cross-section. This holds for all considered scenarios (in particular, irrespectively of whether the scenarios are in the alignment limit or not). 
In scenarios with alignment at the tree level, the inclusion of loop corrections generates a non-zero value of the BSM coupling $\lambda_{hhH}$ and causes the appearance of a resonant contribution that can significantly modify the di-Higgs predictions. 
While the resonant contribution in the alignment limit is suppressed by $\tan\beta$,
even a small deviation from the alignment limit of 1\% is sufficient to induce substantial resonant contributions already at the tree level. 

In conclusion, we demonstrated the sensitivity of the di-Higgs production cross-section, at both the
inclusive and differential levels, to new-physics effects. In this context, the inclusion of the leading NLO
BSM corrections to the process is especially important. The program \anyHH, being
flexible in the choice of model (via the use of \texttt{UFO} model files as inputs) and the definition of a renormalisation scheme, is designed as an
efficient tool for exploring the space of BSM models. This is of importance to provide
a variety 
of phenomenologically relevant scenarios that can serve as input
for experimental searches 
and also to
assess the applicability of experimental results in certain regions of the parameter space of BSM theories.

While this community report mainly focused on the phenomenological examples illustrating the relevance of \anyHH
for the HL-LHC area, a number of additional aspects will be addressed in the future, such as
the discussion of the semi-automatised renormalisation procedure, the impact of interference effects and the technical documentation of the code.

\section*{Acknowledgements}

H.B.\ acknowledges support by the Deutsche Forschungsgemeinschaft (DFG, German Research Foundation) under grant 396021762 – TRR 257: Particle Physics Phenomenology after the Higgs Discovery. J.B.\ is supported by the DFG Emmy Noether Grant No.\ BR 6995/1-1. J.B.\ and G.W.\ acknowledge support by the Deutsche Forschungsgemeinschaft (DFG, German Research Foundation) under Germany's Excellence Strategy --- EXC 2121 ``Quantum Universe'' --- 390833306. This work has been partially funded by the Deutsche Forschungsgemeinschaft (DFG, German Research Foundation) --- 491245950. K.R.\ acknowledges the support of the Spanish Agencia Estatal de Investigaci\'on through the grant IFT Centro de Excelencia Severo Ochoa CEX2020-001007-S funded by MCIN/AEI/10.13039/501100011033. The project that gave rise to these
results received the support of a fellowship from the ``la Caixa'' Foundation (ID 100010434). The fellowship code is LCF/BQ/PI24/12040018.

\bibliography{ref}

\clearpage

\appendix

\end{document}